\documentclass[amsmath,amssymb,aps,pra,10pt,groupedaddress,nofootinbib,longbibliography,superscriptaddress]{revtex4-1}
\usepackage[T1]{fontenc} 
\usepackage[utf8]{inputenc}

\usepackage{graphicx, epsf} 
\usepackage{color} 
\usepackage[table]{xcolor}
\usepackage{dcolumn} 
\usepackage{import} 
\usepackage[skins]{tcolorbox} 
\usepackage{titlesec, setspace} 
\usepackage{listings} 
\usepackage{caption} 
\usepackage[bitstream-charter]{mathdesign} 
\input{Qcircuit}
\usepackage{algorithm}
\usepackage[noend]{algpseudocode}

\usepackage{calrsfs}
\DeclareMathAlphabet{\pazocal}{OMS}{zplm}{m}{n}
\let\mathcal\undefined
\newcommand{\mathcal}[1]{\pazocal{#1}}
\usepackage{courier}

\let\bm\undefined
\newcommand{\bm}[1]{\mathbf{#1}}
\newcommand{\beq}{\begin{equation}}
\newcommand{\eeq}{\end{equation}}

\usepackage[newfloat, frozencache]{minted}

\usepackage{tabularx,booktabs}
\usepackage{cellspace}
\usepackage{placeins}

\newcolumntype{L}[1]{>{\raggedright\arraybackslash}p{#1}}
\arrayrulecolor{white}
\usepackage[hyperfootnotes=true,colorlinks=true,allcolors=blue]{hyperref}

\definecolor{codegreen}{rgb}{0,0.6,0}
\definecolor{codegray}{rgb}{0.5,0.5,0.5}
\definecolor{codepurple}{rgb}{0.58,0,0.82}
\definecolor{backcolour}{RGB}{230, 240, 230}
\definecolor{BlueChill}{RGB}{32, 147, 149}
\definecolor{Shamrock}{RGB}{50, 211, 167}
\titleformat{\section}[display]{\vspace{-1em}}{}{0pt}{\normalfont\sffamily\color{BlueChill}}[\vspace{-3pt}\hrule]
\titleformat{\subsection}[display]{\vspace{-2.5em}}{}{0pt}{\normalfont\sffamily\color{BlueChill}}[\vspace{-1em}]
\titleformat{\subsubsection}[display]{\vspace{-2.5em}}{}{0pt}{\bfseries\color{black}}[\vspace{-1em}]
\makeatletter
\def\frontmatter@abstractfont{\sffamily\color{black}\setstretch{1.5}}%
\def\frontmatter@title@format{\noindent\huge\sffamily\color{BlueChill}}{}%
\def\frontmatter@authorformat{\vspace{1em}\noindent\color{BlueChill}\Large\sffamily}%
\def\frontmatter@affiliationfont{\vspace{1em}\color{BlueChill}\noindent\normalsize\sffamily}%
\def\frontmatter@above@affiliation@script{\vspace{1em}\noindent}%
\def\frontmatter@makefnmark{}
\renewcommand*\frontmatter@date[2][\Dated@name]{\def\@date{}}%
\makeatother

\lstdefinestyle{codeblock}{
  backgroundcolor=\color{backcolour},
  commentstyle=\color{codegreen},
  keywordstyle=\color{blue},
  numberstyle=\tiny\color{codegray},
  stringstyle=\color{codepurple},
  basicstyle=\footnotesize,
  escapechar=\¢,escapebegin=\color{purple}, 
  otherkeywords={with},
  breakatwhitespace=false,
  breaklines=true,
  captionpos=b,
  keepspaces=true,
  language=Python,
  numbers=right,
  numbersep=5pt,
  showspaces=false,
  showstringspaces=false,
  showtabs=false,
  tabsize=2,
  basicstyle=\ttfamily\footnotesize
}
\definecolor{lightgreen}{HTML}{EDF7F4}
\definecolor{xgreen}{HTML}{119A68}
\titleformat{\section}[display]{\vspace{-1em}}{}{0pt}{\Large}[\vspace{-3pt}\color{xgreen}\hrule]
\titleformat{\subsection}[display]{\vspace{-2.5em}}{}{0pt}{\bfseries\color{xgreen}}[\vspace{-1em}]
\titleformat{\subsubsection}[display]{\vspace{-2.5em}}{}{0pt}{\itshape\color{black}}[\vspace{-1em}]

\makeatletter
\def\frontmatter@abstractfont{\color{black}\setstretch{1.5}}%
\def\frontmatter@title@format{\noindent\huge\color{black}}{}%
\def\frontmatter@authorformat{\vspace{1em}\noindent\color{xgreen}\Large}%
\def\frontmatter@affiliationfont{\vspace{1em}\itshape\color{black}\noindent\normalsize}%
\def\frontmatter@above@affiliation@script{\vspace{1em}\noindent}%
\def\frontmatter@makefnmark{}
\renewcommand*\frontmatter@date[2][\Dated@name]{\def\@date{}}%
\makeatother

\usepackage[format=plain,labelfont={bf,it},textfont=it]{caption}
\newenvironment{code}{\captionsetup{type=listing}}{}
\SetupFloatingEnvironment{listing}{name=Codeblock}
\setminted[python]{
	bgcolor=gray!6,
	fontfamily=tt,
	gobble=0,
	fontsize=\small,
	breaklines=true,
	framerule=0.4pt,
	funcnamehighlighting=true,
	tabsize=2,
	obeytabs=false,
	mathescape=false
	samepage=false, 
	showspaces=false,
	showtabs =false,
	texcl=false,
	linenos=true,
	xleftmargin=1pt,
	numbersep=1pt
}

\newmintedfile[pythonfile]{python}{}

\usepackage{silence}
\begin{document}
\title{Differentiable quantum computational chemistry with\\
PennyLane}

\begin{abstract}
This work describes the theoretical foundation for all quantum chemistry functionality in PennyLane, a quantum computing software library specializing in quantum differentiable programming. We provide an overview of fundamental concepts in quantum chemistry, including the basic principles of the Hartree-Fock method. A flagship feature in PennyLane is the differentiable Hartree-Fock solver, allowing users to compute exact gradients of molecular Hamiltonians with respect to nuclear coordinates and basis set parameters. PennyLane provides specialized operations for quantum chemistry, including excitation gates as Givens rotations and templates for quantum chemistry circuits. Moreover, built-in simulators exploit sparse matrix techniques for representing molecular Hamiltonians that lead to fast simulation for quantum chemistry applications. In combination with PennyLane's existing methods for constructing, optimizing, and executing circuits, these methods allow users to implement a wide range of quantum algorithms for quantum chemistry. We discuss how PennyLane can be used to 
implement variational algorithms for calculating ground-state energies, excited-state energies, and energy derivatives, all of which can be differentiated with respect to both circuit and Hamiltonian parameters. We provide an example workflow describing how to jointly optimize circuit parameters, nuclear coordinates, and basis set parameters for quantum chemistry algorithms. We discuss a functionality for reducing the number of qubits by using symmetries and explain how PennyLane can be used to estimate quantum resources needed to implement several quantum algorithms. By combining insights from quantum computing, computational chemistry, and machine learning, PennyLane is the first library for differentiable quantum computational chemistry.

\end{abstract}

\author{Juan Miguel Arrazola}
\affiliation{\textsuperscript{1} Xanadu, 777 Bay Street, Toronto, Canada}
\author{Soran Jahangiri}
\affiliation{\textsuperscript{1} Xanadu, 777 Bay Street, Toronto, Canada}
\author{Alain Delgado}
\affiliation{\textsuperscript{1} Xanadu, 777 Bay Street, Toronto, Canada}
\author{Jack Ceroni}
\affiliation{\textsuperscript{1} Xanadu, 777 Bay Street, Toronto, Canada}
\author{Josh Izaac}
\affiliation{\textsuperscript{1} Xanadu, 777 Bay Street, Toronto, Canada}
\author{Antal Sz\'ava}
\affiliation{\textsuperscript{1} Xanadu, 777 Bay Street, Toronto, Canada}
\author{Utkarsh Azad}
\affiliation{\textsuperscript{1} Xanadu, 777 Bay Street, Toronto, Canada}
\author{Robert A. Lang}
\affiliation{\textsuperscript{2} Chemical Physics Theory Group, Department of Chemistry, University of Toronto, Canada}
\affiliation{\textsuperscript{3} AWS Quantum Technologies, Seattle, Washington 98170, USA}
\author{Zeyue Niu}
\affiliation{\textsuperscript{1} Xanadu, 777 Bay Street, Toronto, Canada}
\author{Olivia Di Matteo}
\affiliation{\textsuperscript{1} Xanadu, 777 Bay Street, Toronto, Canada}
\author{Romain Moyard}
\affiliation{\textsuperscript{1} Xanadu, 777 Bay Street, Toronto, Canada}
\author{Jay Soni}
\affiliation{\textsuperscript{1} Xanadu, 777 Bay Street, Toronto, Canada}
\author{Maria Schuld}
\affiliation{\textsuperscript{1} Xanadu, 777 Bay Street, Toronto, Canada}
\author{Rodrigo A. Vargas-Hern\'andez}
\affiliation{\textsuperscript{2} Chemical Physics Theory Group, Department of Chemistry, University of Toronto, Canada}
\affiliation{\textsuperscript{4} Vector Institute for Artificial Intelligence, Toronto, Canada}
\author{Teresa Tamayo-Mendoza}
\affiliation{\textsuperscript{2} Chemical Physics Theory Group, Department of Chemistry, University of Toronto, Canada}
\affiliation{\textsuperscript{5} Department of Chemistry and Chemical Biology, Harvard University}
\author{Cedric Yen-Yu Lin}
\affiliation{\textsuperscript{3} AWS Quantum Technologies, Seattle, Washington 98170, USA}
\author{Al\'an Aspuru-Guzik}
\affiliation{\textsuperscript{2} Chemical Physics Theory Group, Department of Chemistry, University of Toronto, Canada}
\affiliation{\textsuperscript{4} Vector Institute for Artificial Intelligence, Toronto, Canada}
\affiliation{\textsuperscript{6} Department of Computer Science, University of Toronto, Canada}
\affiliation{\textsuperscript{7} Canadian Institute for Advanced Research (CIFAR) Lebovic Fellow, Toronto, Canada}
\author{Nathan Killoran}
\affiliation{\textsuperscript{1} Xanadu, 777 Bay Street, Toronto, Canada}

\maketitle

\section{Introduction}
Differentiable programming, which is central to machine learning and artificial intelligence, has benefited both from new hardware architectures such as graphical and tensor processing units, and from specialized software libraries for automatic differentiation~\cite{maclaurin2015autograd, abadi2016tensorflow, jax2018github, paszke2019pytorch}. An analogous trend is occurring for quantum computing. Quantum devices across different platforms are now accessible through the cloud, and considerable efforts are ongoing to scale up the number of components and to improve their performance~\cite{reagor2018demonstration, arute2019quantum, wright2019benchmarking, arrazola2021quantum, jurcevic2021demonstration, ebadi2021quantum, pino2021demonstration, alexeev2021quantum, pogorelov2021compact}. Simultaneously, quantum software platforms continue their development with the goal of providing functionality across the entire spectrum of quantum computing~\cite{ibmqiskit, rigettiforest, googlecirq, awsbraket, bergholm2018pennylane, broughton2020tensorflow, killoran2019strawberry, suzuki2020qulacs, fingerhuth2018open}. This includes libraries designed for quantum differentiable programming, with native support for computing gradients of hybrid quantum-classical models~\cite{bergholm2018pennylane, broughton2020tensorflow}.\\

The motivation for building quantum software originates from the underlying interest in quantum computing due to its importance to fundamental science and to its potential to outperform existing approaches, notably for problems related to the simulation of quantum systems~\cite{cirac2012goals, georgescu2014quantum}. Quantum chemistry is recognized as one of the key potential application areas of quantum computing because simulating properties of molecules and materials plays a central role in many industrial processes~\cite{delgado2022simulating, lanyon2010towards, cao2019quantum, bauer2020quantum, mcardle2020quantum}. This has led to considerable interest in developing new and improved quantum algorithms for quantum chemistry, ranging from proof-of-principle methods that can be implemented in simulators and small-scale quantum computers~\cite{peruzzo2014variational,kandala2017hardware, grimsley2019adaptive, mitarai2020theory, cerezo2021variational}, to advanced techniques designed to run on large-scale fault-tolerant machines~\cite{o2016scalable, kivlichan2018quantum, childs2018toward, lee2020even, motta2021low, su2021fault}. Researchers and enthusiasts working in this field have created a demand for quantum software tools designed for quantum chemistry applications, leading to the appearance of software libraries aimed specifically at fulfilling their needs~\cite{mcclean2020openfermion, kottmann2021tequila, stair2021qforte}.\\

This work describes quantum chemistry functionality available in PennyLane, a quantum software library for quantum differentiable programming (\url{www.pennylane.ai})~\cite{bergholm2018pennylane}. By combining methodologies from quantum computing, computational chemistry, and machine learning, PennyLane is the first library built specifically for differentiable quantum computational chemistry. PennyLane users can natively compute gradients of all relevant methods, allowing for a flexible and general approach to the optimization of hybrid quantum-classical algorithms for quantum chemistry. This document is a summary of the theoretical foundation of the software, serving as a technical complement to the online documentation (\url{pennylane.readthedocs.io}). It is intended as a living manuscript to be periodically updated to reflect major new features from all contributing developers.\\

A distinguishing feature of PennyLane's quantum chemistry capabilities is a differentiable implementation of the Hartree-Fock method, as pioneered in Ref.~\cite{tamayo2018automatic} and also implemented in Ref.~\cite{kasim2021dqc}. PennyLane's differentiable solver enables users to compute exact gradients of atomic and molecular orbitals, Hartree-Fock energies, and molecular Hamiltonians, which can then be employed in quantum algorithms. PennyLane provides custom gates and operations for quantum chemistry, including excitation gates as Givens rotations, templates for chemically-inspired circuits, built-in functionality for computing observables, and sparse matrix techniques for representing molecular Hamiltonians. Together with PennyLane's existing methods for designing, optimizing, and executing quantum circuits across simulators and hardware devices, these methods allow users to implement a wide range of quantum algorithms for quantum chemistry, which can be optimized with respect to both circuit and Hamiltonian parameters. PennyLane also includes a collection of tutorials aimed specifically at quantum chemistry, suitable for both beginners and experts.\\

The rest of this manuscript is organized as follows. The first section covers basic principles of quantum chemistry such as the electronic structure problem and methods based on linear combinations of atomic orbitals. We then focus on the differentiable implementation of the Hartree-Fock method, describing fundamental principles of automatic differentiation and explaining how they can be used to construct differentiable implementations of electronic integrals, Fock matrices, and solvers for self-consistent field equations. We also explain how to use these calculations to construct qubit representations of molecular Hamiltonians, which are also fully differentiable.\\

We continue by providing an overview of quantum algorithms for quantum chemistry. We outline the core idea of using quantum computers to build many-body wavefunctions of molecules and compute expectation values of molecular Hamiltonians. We summarize the main methods for achieving this with a focus on variational approaches. The rest of the section describes how to design quantum circuits, compute expectation values, calculate ground and excited-state energies, and obtain exact energy derivatives, which can be used to perform geometry optimization and to compute vibrational normal modes and frequencies of molecules. We describe how to use PennyLane to implement fully differentiable workflows for quantum chemistry capable of simultaneously optimizing gate parameters, nuclear coordinates, and basis set parameters. We also describe how symmetries present in a Hamiltonian can be used to taper off qubits and perform more efficient simulations. Finally, we demonstrate how to estimate quantum resources, such as the number of qubits, gates, and shots, for the variational quantum eigensolver (VQE) and quantum phase estimation (QPE) algorithms. Examples of PennyLane code are used throughout the manuscript as illustrations of the user interface.

\section{Quantum chemistry}\label{Sec:qchem}
This section reviews basic principles of quantum chemistry that are central to the entire manuscript. This includes a discussion of central concepts such as fermionic ladder operators, the occupation-number representation, molecular Hamiltonians, molecular orbitals, and electron integrals. 
 
\subsection{The electronic structure problem}
The underlying laws of quantum mechanics that govern the properties of molecules and materials have been understood for decades, yet the challenge remains to perform efficient and accurate calculations of these properties. A first step in taming the computational challenges of quantum chemistry is to decouple the electronic and nuclear degrees of freedom. This is known as the Born-Oppenheimer approximation, which is supported by the fact that electrons are largely responsible for the chemical properties of molecules, and nuclei are orders-of-magnitude heavier than electrons. The result is the electronic structure problem: solving the time-independent Schr\"{o}dinger equation
\beq
H \ket{\Psi}= E\ket{\Psi},
\eeq
for a collection of interacting electrons in the presence of an electrostatic field generated by the nuclei, which are treated as stationary point particles fixed to their positions. Here $H$ is the electronic Hamiltonian depending parametrically on the positions of the nuclei and $\ket{\Psi}$ is the quantum state of the electronic degrees of freedom. Knowledge of the eigenvalues $E_i$ and eigenstates $\ket{E_i}$ of the Hamiltonian is enough to compute important properties of molecules. Typically it is sufficient to consider only the ground state and the first few excited states.\\

We focus on the second-quantization approach where the antisymmetry of fermionic systems is captured by properties of operators, and states are represented in terms of occupation numbers. More concretely, we consider a set of functions $\{\phi_p(\bm{x})\}$, where $\bm{x}=(\bm{r}, \sigma)$ includes spatial $(\bm{r})$ and spin $(\sigma)$ degrees of freedom. These functions are referred to as spin-orbitals. We introduce the notation
\beq
\ket{n_1, n_2,\ldots, n_M},
\eeq
to denote an electronic state in the space of $M$ spin-orbitals, where $n_p=1$ if $\phi_p$ is occupied by an electron and $n_p=0$ otherwise. Within the restrictions of the chosen set of spin-orbitals, any possible electronic state can be expressed as a superposition of states of this form. \\

To represent general operators, it is useful to define the fermionic ladder operators $a_p, a_p^\dagger$. They satisfy the anticommutation relations
\begin{align}
\{a_p, a_q^\dagger\}&=\delta_{pq},\\
\{a_p, a_q\}&=\{a_p^\dagger, a_q^\dagger\}=0.
\end{align}

Their action on states in the occupation number representation is given by
\begin{align}
a_p\ket{n_1, \ldots, n_M}&=\delta_{1, n_p}(-1)^{N_p}\ket{n_1, \ldots, n_p\oplus 1, \ldots, n_M},\\
a_p^{\dagger}\ket{n_1, \ldots, n_M}&=\delta_{0, n_p}(-1)^{N_p}\ket{n_1, \ldots, n_p\oplus 1, \ldots, n_M},
\end{align} 
where $N_p=\sum_{i=1}^p n_i$ and $\oplus$ denotes addition modulo 2. The phase $(-1)^{N_p}$ applied to the state encodes the exchange symmetry of fermions, meaning that the states themselves do not need to be explicitly antisymmetrized --- an important property that both classical and quantum algorithms benefit from. \\

In the regime where relativistic effects are not important and we can decouple electronic and nuclear degrees of freedom, a molecular Hamiltonian can be represented in terms of these operators as

\beq\label{eq:hamiltonian}
H=\sum_{pq} h_{pq}a_p^\dagger a_q +\frac{1}{2}\sum_{pqrs}h_{pqrs}a_p^\dagger a_q^\dagger a_r a_s.
\eeq
The coefficients $h_{pq}$ and $h_{pqrs}$ are known respectively as the one- and two-electron integrals. They will be described in more detail in future sections.

\subsection{Linear combination of atomic orbitals}

For very simple systems like the hydrogen atom, it is possible to derive analytical solutions of the Schr\"{o}dinger equation. For example, the ground-state wavefunction of the electron in the hydrogen atom is
\beq\label{eq:h_wf}
\Psi(\bm{r})=\frac{1}{\sqrt{\pi}}e^{-r},
\eeq
where $\bm{r}=(x,y,z)$ is the electron coordinate vector, $r= \sqrt{x^2+y^2+z^2}$, and we use atomic units that set the Bohr radius equal to one. Analytical solutions for more complicated systems quickly become intractable, requiring the use of numerical methods. Any function can be represented in terms of a complete basis, for example the set of all plane waves of the form $e^{-i\bm{k}\cdot \bm{r}}$. Since exact representations require an infinitely large basis set, it is customary to define a restricted basis set that can be used as an approximation to the continuum limit. The challenge is to find basis sets containing few elements that can still provide accurate representations.\\

The most common approach in quantum chemistry is to introduce localized basis sets that are appropriate for describing functions centered at the nuclei of each atom. These are used to represent localized functions known as atomic orbitals. By taking linear combinations of atomic orbitals, it is possible to represent molecular orbitals that describe how electrons are distributed around the entire molecule. This strategy is known as the linear combination of atomic orbitals (LCAO)~\cite{lehtola2020overview}.\\

Atomic orbitals $\chi(\bm{r})$ are expanded in terms of primitive functions $\psi(\bm{r})$ as
\beq
\chi(\bm{r})=\mathcal{N}\sum_{i=1}^n a_i \psi_i(\bm{r}, \alpha_i),
\eeq
where the $a_i$ are coefficients, $\alpha_i$ denote possible internal parameters of the primitive functions, and $\mathcal{N}$ is a normalization constant. Here we are implicitly assuming that these functions are centered at the origin, but their centers can be translated to the nuclear coordinates $\bm{R}$ by setting $\bm{r}\rightarrow \bm{r-R}$. This choice is a key feature of the LCAO approach: electron wave functions have cusps centered at the nuclei where the Coulomb potential peaks, so it is useful to employ nuclei-centered orbitals that can replicate this behaviour.  Molecular orbitals are denoted by $\phi(\bm{r})$ and can be expressed as
\beq
\phi(\bm{r})=\sum_{i} C_i\chi_i(\bm{r}),
\eeq
where the $C_i$ are real coefficients and the index $i$ runs over all atomic orbitals.  
Molecular orbitals can thus be viewed as a double expansion: they are a linear combination of atomic orbitals, which are themselves linear combinations of primitive basis set functions. In the restricted case, particles of different spin occupy the same spatial orbitals, allowing us to drop the spin label and consider both types of spin-orbitals on the same footing.\\

Experts have worked for decades to develop a vast array of increasingly sophisticated basis sets for chemistry applications~\cite{nagy2017basis, pritchard2019new}. The coefficients $a_i$, which have been optimized to describe the electronic structure of isolated atoms, are typically used regardless of the molecule containing the atoms. The parameters that are individually optimized for each molecule are the coefficients $C_i$. Large basis sets are needed for highly-accurate simulations, which leads to increased computational costs. This establishes a fundamental trade-off between accuracy and efficiency which is particularly important for quantum algorithms, where the number of qubits typically increases with the number of elements in the basis set.  \\

The one- and two-electron integrals $h_{pq}$, $h_{pqrs}$ introduced in Eq.~\eqref{eq:hamiltonian} can be interpreted as matrix elements of the Hamiltonian in the basis of molecular orbitals. Their full expressions are
\begin{align}
h_{pq}&=\int d\bm{x} \,\phi_p^*(\bm{x})\left(-\frac{\nabla^2}{2}-\sum_{i=1}^N\frac{Z_i}{|\bm{r}-\bm{R}_i|}\right)\phi_q(\bm{x}),\label{eq:h_pq}\\
h_{pqrs}&=\int d\bm{x}_1 d\bm{x}_2\, \frac{\phi_p^*(\bm{x}_1)\phi_q^*(\bm{x}_2)\phi_r(\bm{x}_2)\phi_s(\bm{x}_1)}{|\bm{r}_1-\bm{r}_2|},\label{eq:h_pqrs}
\end{align} 
where $Z_i$ is the atomic number of the $i$-th nucleus and the sum is taken over all atoms in the molecule. The one-electron integral contains contributions from the electronic kinetic energy and the electron-nuclear attraction, while the two-electron integral represents energy contributions from the Coulomb repulsion between pairs of electrons. \\

Molecular Hamiltonians are defined in terms of the integrals in Eqs.~\eqref{eq:h_pq} and~\eqref{eq:h_pqrs}, which depend explicitly on the choice of molecular spin-orbitals. In the following section, we discuss the Hartree-Fock method, which can be employed to obtain optimized spin-orbitals. In PennyLane, the Hartree-Fock method is implemented in a fully differentiable manner, allowing users to compute gradients of all relevant quantities with respect to basis set parameters and nuclear coordinates. This is achieved using the techniques of automatic differentiation, which we describe briefly in comparison to other strategies such as symbolic and numerical differentiation.

\section{Differentiable Hartree-Fock}

The ability to compute derivatives of functions is a central task in science and mathematics. Symbolic differentiation obtains derivatives of an input function by direct mathematical manipulation, for example using standard strategies of differential calculus. These can be performed by hand or with the help of computer algebra software. The resulting expressions are exact, but the symbolic approach is of limited scope, particularly since many functions are not known in explicit analytical form. Symbolic methods also suffer from the expression swell problem where careless usage can lead to exponentially large symbolic expressions. Numerical differentiation is a versatile but unstable method, often relying on finite differences to calculate approximate derivatives. This is problematic especially for large computations consisting of many differentiable parameters~\cite{baydin2018automatic}.\\

Automatic differentiation is a computational strategy where a function implemented using computer code is differentiated by expressing it in terms of elementary operations for which derivatives are known~\cite{griewank1989automatic, baydin2018automatic}. The gradient of the target function is then obtained by applying the chain rule through the entire code. In principle, automatic differentiation can be used to calculate derivatives of a function using resources comparable to those required to evaluate the function itself. This strategy has gained significant attention in machine learning, and consequently several software packages for automatic differentiation have been developed~\cite{maclaurin2015autograd, abadi2016tensorflow, jax2018github, paszke2019pytorch}.\\

In this section, we describe the Hartree-Fock method for quantum chemistry and outline its implementation in PennyLane using principles of automatic differentiation. Building upon work pioneered in Ref.~\cite{tamayo2018automatic}, PennyLane provides built-in methods for constructing atomic and molecular orbitals, building Fock matrices, and solving the self-consistent Hartree-Fock equations to obtain optimized orbitals, which can be used to construct molecular Hamiltonians. PennyLane allows users to natively compute derivatives of all these objects with respect to the underlying parameters. The implementation makes use of differentiable transforms, allowing the entire workflow to be differentiable while maintaining a simple and flexible user interface. All physical quantities in this context are defined in atomic units. \\

The main goal of the Hartree-Fock method is to obtain molecular spin-orbitals that minimize the energy of a state where electrons are treated as independent particles occupying the lowest-energy orbitals. Here we provide an overview of the main steps of this process and their PennyLane implementation. An in-depth pedagogical review can be found in Ref.~\cite{lehtola2020overview}. We focus on the restricted Hartree-Fock method, where each spatial orbital is occupied by two electrons with opposite spin.

\subsection{Basis sets, orbitals, and molecules}
The design and choice of basis sets is largely guided by their usefulness in computational methods. Part of the challenge in optimizing molecular orbitals, building Hamiltonians, and performing electronic structure calculations is to compute integrals over primitive basis functions. Most modern methods in quantum chemistry employ Gaussian functions as the primitive basis because they are simple to evaluate analytically and numerically~\cite{fermann2020fundamentals}. These are functions of the form
\beq
\psi(\bm{r}, \alpha)=x^ly^mz^ne^{-\alpha r^2},
\eeq
where for simplicity we assume are centered at the origin. Gaussian functions $e^{-\alpha r^2}$ are smoother at the origin and decay more quickly than Slater-type functions $e^{-\alpha r}$, as in Eq.~\eqref{eq:h_wf}. Nonetheless, linear combinations of Gaussians can be used to approximate Slater-type orbitals. This is the strategy of the STO-$n$G class (Slater-type orbitals with $n$ primitive Gaussians), of which the simplest example that is commonly used is the STO-3G basis set. This consists of atomic orbitals of the form
\beq
\chi(\bm{r})=\sum_{i=1}^3 a_i\psi(\bm{r}, \alpha_i),
\eeq 
where the contraction coefficients $a_i$ and the exponents $\alpha_i$ are typically predefined for each atom, but in principle can be optimized for specific molecules. PennyLane provides native support for the STO-3G, 6-31G, 6-311G and cc-pVDZ basis sets in the differentiable solver, while also allowing users to import other basis sets. Moreover, PennyLane provides native support for differentiating atomic orbitals with respect to contraction coefficients, exponents, and nuclear coordinates. \\

The \texttt{Molecule} class can be used to encapsulate all relevant information about a system. By specifying a list of atomic symbols and nuclear coordinates, we can instantiate a molecule object containing as attributes the charge of the molecule, number of electrons, nuclear charges, number of orbitals, Gaussian exponents $\alpha$, Gaussian centers, contraction coefficients $a$, and angular momentum values $(l,m,n)$ of the basis functions, which can also be optionally passed by the user. This enables sophisticated functions that depend on all these parameters to simply take a molecule object as an argument.

\begin{code}
	\pythonfile{code/molecule.py}
	\caption{Constructing a molecule object for the hydrogen molecule.}
	\label{code:molecule}
\end{code}

\subsection{Matrices and integrals}
To group all coefficients in a single object, we can define the coefficient matrix $\bm{C}$ with entries $C_{\mu i}$ such that the $i$-th molecular orbital can be expressed as
\beq
\phi_i(\bm{r})=\sum_{\mu} C_{\mu i}\chi_\mu(\bm{r}),
\eeq 
where Greek letters are used to denote sums over basis functions. We also define the matrix $\bm{P}$ with entries
\beq
P_{\mu\nu}=\sum_{i}C_{\mu i}C_{\nu i},
\eeq
which leads to a concise expression for the Hartree-Fock energy
\beq
E = 2 \sum_{\mu\nu}P_{\mu\nu}H_{\mu\nu}+ \sum_{\mu\nu} P_{\mu\nu} \left (2 J_{\mu\nu}-K_{\mu\nu}  \right ).
\eeq
The matrices appearing in this expansion are the core Hamiltonian $\bm{H}$ with entries
\beq\label{eq:core}
H_{\mu\nu} = \int dr \,\chi^*_\mu(\bm{r})\left(-\frac12 \nabla^2 - \sum_{i=1}^N \frac{Z_i}{|\bm{r}-\bm{R_i}|}\right)\chi_\nu(\bm{r}),
\eeq
the Coulomb matrix $\bm{J}$ with entries
\beq
J_{\mu\nu}=\sum_{\eta\gamma}\,P_{\eta\gamma}(\mu\nu|\eta\gamma),
\eeq
and the exchange matrix $\bm{K}$ with entries
\beq
K_{\mu\nu}=\frac12\sum_{\eta\gamma}\,P_{\mu\nu}(\mu\eta|\nu\gamma),
\eeq
where the electron repulsion integral is defined as
\beq\label{eq:eri}
(\mu\nu|\eta\gamma)=\int d\bm{r}_1\, d\bm{r}_2\,\frac{ \chi^*_\mu(\bm{r}_1)\chi^*_\nu(\bm{r}_2)\chi_\eta(\bm{r}_1)\chi_\gamma(\bm{r}_2)}{|\bm{r}_1-\bm{r}_2|}.
\eeq
The core Hamiltonian does not depend on the coefficients $C_{\mu i}$, but the total energy depends on them through the matrix $\bm{P}$.\\  

The goal of the Hartree-Fock method is to find the coefficients $C_{\mu i}$ that minimize the total energy function. While this problem could be tackled through different strategies, a closed-form solution is known that can be used to directly obtain the optimized molecular orbitals~\cite{fermann2020fundamentals}. Central to this approach is the Fock matrix $\bm{F}$, whose entries are defined as

\beq
F_{\mu\nu}:= \frac{\partial E}{\partial P_{\mu\nu}}=H_{\mu\nu} + 2 J_{\mu\nu}- K_{\mu\nu}.
\eeq

PennyLane provides functionality to compute core Hamiltonians, Coulomb matrices, exchange matrices, repulsion integrals, Fock matrices, and total energy, all of which are differentiable with respect to nuclear coordinates, contraction coefficients, and Gaussian exponents. By expanding the atomic orbitals in terms of the primitive Gaussian functions, the core Hamiltonian integrals of Eq.~\eqref{eq:core} and the electron repulsion integrals of Eq.~\eqref{eq:eri} can be expressed as linear combinations of Gaussian integrals, for which there exist closed-form analytical formulas~\cite{helgaker1995gaussian}. This enables an implementation that is directly compatible with automatic differentiation techniques. PennyLane currently builds on Autograd~\cite{maclaurin2015autograd} as a lightweight approach to computing derivatives.

\subsection{Self-consistent equations}
The problem of finding the optimal coefficients $C_{\mu i}$ can be reduced to solving the Roothaan-Hall equation~\cite{pople1954self, pople1992kohn, lehtola2020overview}
\beq\label{eq:RH}
\bm{FC}=\bm{SCE},
\eeq
where $\bm{F}$ and $\bm{C}$ are respectively the Fock and coefficient matrix, $\bm{E}$ is a diagonal matrix of eigenvalues, and $\bm{S}$ is the overlap matrix with entries
\beq
S_{\mu\nu}=\int d\bm{r}\, \chi_\mu(\bm{r})^*\chi_\nu(\bm{r}).
\eeq
Equation~\eqref{eq:RH} can be further simplified by transforming the Fock and coefficient matrices. Diagonalizing the overlap matrix as
\beq
\bm{S}=\bm{V}\,\bm{D}\, \bm{V}^T,
\eeq
where $\bm{D}$ is a diagonal matrix, we define $\bm{X}=\bm{V}\,\bm{D}^{-1/2}\, \bm{V}^T$, $\bm{C}=\bm{X}\,\tilde{\bm{C}}$, and $\tilde{\bm{F}}=\bm{X}^T\bm{F}\,\bm{X}.$ The equation can then be cast in its standard form
\beq\label{eq:RH-standard}
\tilde{\bm{F}}\tilde{\bm{C}}=\tilde{\bm{C}}\bm{E}.
\eeq
This is an eigenvalue problem that can be solved using standard techniques in linear algebra, which are compatible with automatic differentiation.\\

The Fock matrix depends implicitly on the coefficient matrix. By setting an initial value for $\tilde{\bm{C}}_0$, computing the resulting matrix $\tilde{\bm{F}}_0$, and solving Eq.~\eqref{eq:RH-standard}, it is possible to end with a solution $\tilde{\bm{C}}_1$ that is inconsistent with the original choice of $\tilde{\bm{C}}_0$. Instead, we seek a self-consistent solution by iteratively repeating the process until $\bm{C}^k=\bm{C}^{k+1}$. Once a self-consistent solution is found, the resulting coefficients can be used to construct optimized molecular orbitals, compute the Hartree-Fock energy, and build molecular Hamiltonians. Different initialization strategies are possible, but a simple and effective approach is to set the first coefficient matrix equal to zero, meaning the only contribution to the energy arises from the core Hamiltonian. The example code below shows how to calculate the Hartree-Fock energy and its gradient with respect to nuclear coordinates for the simple case of the hydrogen molecule, which we focus on for subsequent code examples:

\begin{code}
	\pythonfile{code/hf.py}
	\caption{Calculating the Hartree-Fock energy and its derivative with respect to nuclear coordinates. This uses the same quantities defined in Codeblock~\ref{code:molecule}. The nuclear coordinates that define the molecular geometry are differentiable by default.}
	\label{code:hf}
\end{code}

This simple user interface for constructing molecule objects makes use of default values for the basis set, charge, multiplicity, and active space of the molecule. Advanced users can also specify custom values for all these quantities.

\section{Qubit Hamiltonians}
The starting point of many quantum algorithms for quantum chemistry is the Hamiltonian describing the system of interest. The molecule itself is described by the set of its constituent atoms and their coordinates in three-dimensional space. Within a given basis set, this information is sufficient to build the Fock matrix and implement the Hartree-Fock method. The resulting optimized orbitals are then used to compute the one- and two-electron integrals that specify the molecular Hamiltonian in terms of fermionic ladder operators. \\

To execute a quantum algorithm, it is necessary to express the Hamiltonian as a qubit operator while ensuring that it retains all the required fermionic exchange symmetries. This in turn requires a description of fermionic states in terms of qubit states. These transformations are performed using fermionic-to-qubit mappings~\cite{mcardle2020quantum}. The most widely used is the Jordan-Wigner transformation~\cite{jordan1928paulische}. Here, a state $|n_1,\ldots, n_M\rangle$ in the occupation number representation on $M$ spin-orbitals can be directly mapped to a state $|n_1\rangle|n_2\rangle\cdots |n_M\rangle$ on $M$ qubits, where the state of the $i$-th qubit is $\ket{1} $ if the corresponding spin-orbital is occupied, and $\ket{0}$ otherwise. Fermionic ladder operators are transformed as
\begin{align}
a_p = \frac{1}{2}Z_0\otimes \cdots Z_{p-1} \otimes(X_p+iY_p),\\
a_p^{\dagger} = \frac{1}{2}Z_0\otimes \cdots Z_{p-1} \otimes(X_p-iY_p),
\end{align}  
where $X,Y,Z$ are Pauli matrices. This results in a Hamiltonian of the form
\beq
H = \sum_j h_j P_j,
\eeq
where $P_j$ is a tensor product of the Pauli matrices $I,X,Y,Z$. In PennyLane, a qubit Hamiltonian can be easily obtained with the \texttt{molecular\_hamiltonian()} function:

\begin{code}
	\pythonfile{code/hamiltonian.py}
	\caption{Constructing the Hamiltonian for the hydrogen molecule, using the same definitions as in Codeblock~\ref{code:molecule}. }
	\label{code:hamiltonian}
\end{code}

As an alternative to the differentiable Hartree-Fock solver, PennyLane can also interface with the external packages PySCF~\cite{sun2018pyscf} or Psi4~\cite{turney2012psi4} to calculate the one- and two-electron integrals and use functionality in OpenFermion~\cite{mcclean2020openfermion} to construct the qubit Hamiltonian. Users can also construct OpenFermion Hamiltonians themselves and map them to PennyLane Hamiltonians.\\

The \texttt{molecular\_hamiltonian()} function can take differentiable parameters as input to construct the Hamiltonian. Under the hood, this function uses the differentiable solver to compute the optimized Hartree-Fock orbitals, calculate the one- and two-electron integrals, and perform the Jordan-Wigner fermionic-to-qubit mapping to construct a PennyLane Hamiltonian. \\

Differentiable qubit Hamiltonians can be constructed by defining the atomic symbols and nuclear coordinates for a given molecule and specifying the molecular parameters that will be differentiated. Currently, the exponents, contraction coefficients, and centers of the Gaussian basis functions can be specified to be differentiable by setting \texttt{requires\_grad=True} when their initial values are defined. The following example code shows how to build a differentiable qubit Hamiltonian when the atomic coordinates and the basis set parameters are all differentiable:

\begin{code}
	\pythonfile{code/hamiltonian_diff.py}
	\caption{Constructing the differentiable Hamiltonian for the hydrogen molecule.}
	\label{code:hamiltonian_diff}
\end{code}

PennyLane also provides functionality to construct sparse matrix representations of molecular Hamiltonians. In simulators, this is used by PennyLane to perform fast computations of expectation values. PennyLane also allows users to directly diagonalize the target Hamiltonian from its sparse-matrix representation, which can be used to benchmark the performance of quantum algorithms without the need to perform additional quantum chemistry calculations, such as full-configuration interaction.
Besides molecular Hamiltonians, PennyLane also provides support for constructing other observables, which currently require interfacing with external packages. These include total particle number, total spin, spin projection, dipole moment, and user-specified observables.

\section{Quantum circuits}

Quantum algorithms for quantum chemistry construct states of many-body fermionic systems and extract relevant information from them. The most popular methods to achieve this are quantum phase estimation and variational algorithms. In quantum phase estimation~\cite{nielsen2010quantum}, the eigenvalues of a Hamiltonian $H$ are encoded in the eigenvalues of a unitary $U$, for example through the time-evolution operator $U=e^{-iHt}$. The algorithm samples from the spectrum of $U$ by performing repeated calls to an oracle implementing $U$ controlled on the state of auxiliary qubits. The auxiliary qubits are then rotated using an inverse quantum Fourier transform and measured to reveal a sampled eigenvalue. If the input state has a sufficiently large overlap with a target eigenstate of $H$, quantum phase estimation can probabilistically prepare the target eigenstate and compute its corresponding eigenvalue. \\

This is a remarkable capability that underlies the long-term potential of quantum computing for quantum chemistry. However, quantum phase estimation requires a large number of qubits and long-depth circuits that make it challenging to implement in both hardware and simulators, even for small systems. Still, PennyLane provides a general template for quantum phase estimation that can be leveraged by advanced users to study prototype applications to quantum chemistry.\\

In variational algorithms, the goal is to design quantum circuits that can be optimized to prepare fermionic states of interest. This allows for shorter-depth circuits without the need for auxiliary qubits, making them more appropriate for simulation and implementation in noisy hardware. However, variational algorithms lack the theoretical guarantees of quantum phase estimation and struggle to compute expectation values efficiently when implemented in hardware~\cite{gonthier2020identifying}. Optimization can often be carried out using gradient-based methods, meaning variational algorithms are amenable to the methodologies of quantum differentiable programming that are PennyLane's core strength. Below we discuss the various strategies that can be used in PennyLane to create, evaluate, and optimize variational quantum circuits for quantum chemistry.

\subsection{Givens rotations}
Under the Jordan-Wigner transformation, a qubit is associated with each spin-orbital and its states are used to represent occupied $\ket{1}$ or unoccupied $\ket{0}$ spin-orbitals. When designing quantum circuits for quantum chemistry, it is useful to employ only particle-number conserving gates guaranteeing that the trial space is within the correct particle number subspace. The simplest non-trivial particle-conserving gate is a two-qubit transformation that couples the states $\ket{01}, \ket{10}$ as
\begin{align}
U\ket{01} = a\ket{01} + b \ket{10},\\
U\ket{10} = c\ket{10} + d \ket{01},
\end{align}
while acting as the identity on $\ket{00}$ and $\ket{11}$. Here $|a|^2+|c|^2=|b|^2+|d|^2=1$ and $ab^*+cd^*=0$ to ensure unitarity. This is an example of a Givens rotation: a two-dimensional rotation in the subspace of a larger Hilbert space. Restricting to the case of real parameters, this gate can be written as
\beq
G(\theta)=\begin{pmatrix}
1 & 0 & 0 & 0\\
0 & \cos (\theta/2) & -\sin (\theta/2) & 0\\
0 & \sin(\theta/2) & \cos(\theta/2) & 0\\
0 & 0 & 0 & 1
\end{pmatrix},\label{eq:givens}
\eeq
where we use the ordering $\ket{00}, \ket{01}, \ket{10}, \ket{11}$ of computational basis states. We call this a single-excitation gate 
because the states $\ket{10}, \ket{01}$ differ by the excitation of a single particle. Similarly, we can consider the four-qubit gate that couples the states $\ket{1100}$ and $\ket{0011}$ as
\begin{align}
G^{(2)}(\theta)\ket{0011} = \cos (\theta/2)\ket{0011} + \sin (\theta/2) \ket{1100},\\
G^{(2)}(\theta)\ket{1100} = \cos (\theta/2)\ket{1100} - \sin (\theta/2) \ket{0011}.
\end{align}
This is an example of a double-excitation gate. Single-excitation and double-excitation gates are implemented natively in PennyLane, together with decompositions into elementary gate sets and analytical gradient rules~\cite{kottmann2021feasible, wierichs2021general}.\\

It was proven in Ref.~\cite{arrazola2021universal} that controlled single-excitation gates are universal for particle-conserving unitaries, establishing the role of Givens rotations as the ideal building blocks of quantum circuits for quantum chemistry. PennyLane places an emphasis on single and double-excitation gates in the construction of variational circuits, as shown in the example code below. Using the \texttt{qnode} decorator, quantum circuits can be connected to a specific device performing its execution, whether hardware or simulators. In this case the circuit is executed by the built-in \texttt{default.qubit} simulator, which returns the expectation value of the Hamiltonian:

\begin{code}
	\pythonfile{code/excitation_circuit.py}
	\caption{An example circuit built using excitation gates based on Givens rotations. The circuit is initialized in the Hartree-Fock state for a system with two electrons in two spin-orbitals. It then applies a double-excitation gate and two single-excitation gates that are chosen to preserve spin.}
	\label{code:excitations}
\end{code}

\subsection{Templates}
A common approach in the design of quantum circuits for quantum chemistry is to propose a fixed circuit architecture that can be used for a variety of molecules. Also known as ans\"atze, these circuits aim to provide a general strategy for creating states of interest in quantum chemistry. Circuits composed of multiple gates can be pre-coded in PennyLane as templates, which can then be called in the same way as individual operations. This allows for a simple interface to build circuits that implement various ans\"atze for quantum chemistry.\\

PennyLane currently provides templates for the unitary coupled-cluster singles and doubles (UCCSD)~\cite{romero2018strategies, anand2021quantum}, particle-conserving gate fabrics~\cite{anselmetti2021local}, k-UpCCGSD~\cite{lee2018generalized}, and the AllSinglesDoubles~\cite{arrazola2021universal} ansatz consisting of all spin-conserving single and double-excitation gates, which is shown in the example code below:

\begin{code}
	\pythonfile{code/template.py}
	\caption{Using a template to implement a circuit consisting of all spin-conserving single- and double-excitation gates acting on the Hartree-Fock state defined on a system with two electrons and four spin-orbitals, as in the hydrogen molecule in a minimal basis set.}	\label{code:template}
\end{code}

\subsection{Adaptive circuits}
A fixed circuit ansatz may work well in many cases but it will not be optimized for the specific problem at hand. This motivates strategies that select gates using information from the system being studied, which can lead to more efficient circuits. The most common approach for adaptively creating circuits is to select gates that have a large gradient with respect to the relevant cost function. By design, PennyLane allows the computation of gradients with respect to all parametrized gates in a quantum circuit.\\

The strategy of the ADAPT-VQE algorithm~\cite{grimsley2019adaptive, tang2019qubit} is to define a pool of gates, apply each to an initial state (typically the Hartree-Fock state), and select the gate with the largest gradient. This gate is then optimized and the resulting gate parameter fixed, which defines a new initial state. The procedure is repeated until a convergence criterion is reached. The \texttt{AdaptiveOptimizer()}  functionality in PennyLane can be used to implement algorithms such as ADAPT-VQE:

\begin{code}
	\pythonfile{code/adapt.py}
	\caption{Building a circuit adaptively by selecting and adding gates from a user-defined collection of operators. The initial state is defined on a system with two electrons and four spin-orbitals, as in the hydrogen molecule in a minimal basis set.}	\label{code:adapt}
\end{code}

\section{Ground and excited-state energies}
Once a quantum circuit has been defined, the goal of variational algorithms is to optimize the circuit with respect to an appropriate cost function. As discussed before, it is often sufficient to compute ground and excited-state energies of molecular Hamiltonians. This section discusses algorithms for performing these calculations and their PennyLane implementation.

\subsection{Ground states}

The variational quantum eigensolver (VQE) is an algorithm that computes approximate ground-state energies by optimizing the parameters of a quantum circuit with respect to the expectation value of a molecular Hamiltonian. More precisely, using $\theta=(\theta_1,\ldots, \theta_n)$ to denote the parameters of a variational circuit, $\ket{\psi(\theta)}$ to denote the state prepared by the circuit, and $H$ for the molecular Hamiltonian, the goal of VQE is to optimize the quantum circuit in order to minimize the cost function
\beq
C(\theta) = \bra{\psi(\theta)}H\ket{\psi(\theta)}.
\eeq

This is a broad approach that encompasses a variety of methods for circuit design and parameter optimization. As discussed above, PennyLane provides functionality to build molecular Hamiltonians and design quantum circuits using individual gates, templates, or adaptive methods. To perform the circuit optimization, users have access to a wide array of gradient-based optimizers that leverage PennyLane's built-in ability to compute gradients of quantum circuits. This includes quantum-aware optimizers such as the quantum natural gradient~\cite{stokes2020quantum}, Rotosolve~\cite{vidal2018calculus, ostaszewski2021structure, wierichs2021general}, and Rotoselect~\cite{ostaszewski2021structure}. \\

To compute expectation values of Hamiltonians using simulators, PennyLane converts the Hamiltonian to a sparse matrix, then uses the vector representation of the state to compute the expectation value $\langle H\rangle = \bra{\psi}H\ket{\psi}$ directly using matrix-vector multiplication. It is also possible to directly pass a sparse matrix representation of the Hamiltonian to PennyLane's \texttt{expval()} function. Gradients are obtained using parameter-shift rules~\cite{schuld2019evaluating}, backpropagation, or the adjoint method. \\

To calculate expectation values in hardware, it is necessary to compute the expectation of each term in the Hamiltonian's expansion as a linear combination of Pauli words
\beq
\langle H\rangle = \sum_jh_j \langle P_j\rangle.
\eeq
The expectation values $\langle P_j\rangle$ can be calculated by performing only additional single-qubit rotations because the Pauli words $P_j$ are tensor products of local qubit operators. One major challenge is the large number of terms in this expansion, which increases rapidly for larger molecules. This leads to the requirement for a total number of samples that scales dramatically with system size; a major obstacle for scaling these algorithms. To partially alleviate this problem, it is possible to group Pauli words into sets of mutually-commuting operators whose expectation values can then be calculated from the same measurement statistics. This functionality is provided in PennyLane's \texttt{grouping} module, and can be activated by passing the keyword argument \texttt{optimize=True} when computing the expectation value.\\

The code below shows a full VQE workflow in PennyLane for computing the ground state of the hydrogen molecule in the default minimal basis set.

\begin{code}
	\pythonfile{code/vqe.py}\label{code:vqe}
	\caption{VQE workflow to compute the ground-state energy of the hydrogen molecule.} \label{code:vqe}
\end{code}

By performing ground-state calculations for Hamiltonians $H(\bm{R})$ defined on different nuclear coordinates $\bm{R}$, it is possible to compute ground-state potential energy surfaces of molecules, which are defined by the energy function
\beq\label{Eq:energy}
    E(\bm{R}) = \min_{\ket{\psi}} \bra{\psi}H(\bm{R})\ket{\psi} = \bra{\psi_0(\bm{R})}H(\bm{R})\ket{\psi_0(\bm{R})},
\eeq
where $\ket{\psi_0(\bm{R})}$ is the ground state, which depends on the nuclear coordinates $\bm{R}$. The minima of a potential energy surface correspond to equilibrium geometries, while local maxima correspond to transition states. Computing the energies at these configurations allows us to calculate activation energies and chemical reaction rates.

\subsection{Excited states}
The simplest method to extend ground-state algorithms to excited states is to leverage the fact that eigenstates are mutually orthogonal~\cite{jones2019variational, higgott2019variational}. Excited-state energies can be computed by adding penalty terms to the cost function that enforce orthogonality with lower-lying eigenstates. For example, to optimize a circuit that prepares the first excited state we employ the cost function
\begin{align}
C^{(1)}(\theta)&=\bra{\psi(\theta)}H^{(1)}\ket{\psi(\theta)},\\
H^{(1)}&=H+\beta \ket{\psi_0}\bra{\psi_0},
\end{align} 
where $\beta>0$ is an adjustable penalty parameter. The ground state $\ket{\psi_0}$ is also an eigenstate of $H^{(1)}$ with eigenvalue $E_0+\beta$, where $E_0$ is the ground-state energy of $H$. Setting $\beta>E_1-E_0$ ensures that the lowest-energy eigenstate of $H^{(1)}$ is actually the first excited state of $H$. This procedure can be iterated to find the $k$-th excited state by minimizing the cost function
\begin{align}
C^{(k)}(\theta)&=\bra{\psi(\theta)}H^{(k)}\ket{\psi(\theta)},\\
H^{(k)}&=H+\sum_{i=0}^{k-1}\beta_i \ket{\psi_i}\bra{\psi_i}.
\end{align}
PennyLane users can perform excited-state calculations using the same functionalities as for ground-state computations by adjusting the cost function accordingly. It is also possible to employ the built-in spin and dipole moment observables to add penalties that select excited states corresponding to desired transitions.\\

These extended cost functions can be readily evaluated using simulators, but require additional resources in hardware. Penalty terms of the form $|\bra{\psi(\theta)}\psi_i\rangle|^2$ can be computed using a SWAP test, which requires roughly doubling the number of qubits for the same circuit depth. An alternative is to apply a unitary $V^\dagger$ at the end of the circuit, where $V\ket{0}=\ket{\psi_i}$. The desired overlap can be computed as $|\langle\psi(\theta)|\psi_i\rangle|^2=|\bra{0}U(\theta)V^\dagger\ket{0}|^2$, which is the probability of measuring the all-zero state in the adjusted circuit and $U(\theta)$ is a unitary preparing the trial ground state. This maintains the same number of qubits while roughly doubling the circuit depth.

\section{Energy derivatives}
Ground and excited-state energies provide valuable information about the properties of molecules, but they are not exhaustive in revealing all relevant information. Further insights can be obtained by studying gradients of the total energy. This provides the ability to calculate forces on nuclei, perform molecular geometry optimization, and compute vibrational normal modes and frequencies in the harmonic approximation. This section describes the theory of energy derivatives in quantum chemistry and explains how to implement related algorithms in PennyLane.

\subsection{Nuclear forces and geometry optimization}

The force experienced by the nuclei of each atom in a molecule is given by the gradient of the total energy with respect to the nuclear coordinates
\beq
F(\bm{R}) = -\nabla_{\bm{R}} E(\bm{R}),
\eeq
where $E(\bm{R})$ is defined as in Eq.~\eqref{Eq:energy}. From the Hellmann-Feynman theorem, the derivative of the ground-state energy with respect to the coordinates of the $i$-th atom can be written as

\beq\label{Eq:energy_gradient}
    \frac{\partial E(\bm{R})}{\partial R_i} = \Big \langle\psi_0(\bm{R})\Big |\frac{\partial H(\bm{R})}{\partial R_i}\Big | \psi_0(\bm{R})\Big\rangle.
\eeq

This is an analytical expression that can be calculated from a circuit that has been optimized to prepare $\ket{\psi_0(\bm{R})}$. Crucially, it requires a method to compute the Hamiltonian derivatives $\partial H(\bm{R})/\partial R_i$. One of the key advantages of PennyLane's differentiable Hartree-Fock solver is that energy can be calculated directly and exactly using methods of automatic differentiation, as illustrated in the example code below:

\begin{code}
	\pythonfile{code/forces.py}
	\caption{The nuclear forces can be computed exactly using the differentiable Hartree-Fock solver to compute gradients of an optimized circuit that calculates the ground-state energy.}
	\label{code:forces}
\end{code}

More broadly, as described in Ref.~\cite{delgado2021variational}, Hamiltonian derivatives permit the study of cost functions that depend both on circuit parameters $\theta$ and Hamiltonian parameters such as the nuclear coordinates $\bm{R}$:

\beq
C(\theta, \bm{R}) = \bra{\psi(\theta)}H(\bm{R})\ket{\psi(\theta)}.
\eeq
PennyLane users can compute gradients of this cost function with respect to both set of parameters, allowing for a joint optimization of the cost function. The results are optimal circuit parameters $\theta^*$ that prepare an approximate ground state and optimal nuclear coordinates $\bm{R}^*$ that describe the equilibrium geometry of the molecule.

\subsection{Hessians and vibrational modes}

Expressions for higher-order energy derivatives can also be obtained~\cite{mitarai2020theory, azadsingh2021}, but in this case there are contributions arising from derivatives of
the ground state $\ket{\psi_0(\bm{R})}$, which themselves depend on derivatives of the optimal parameters $\theta^*(\bm{R})$. The second-order
derivatives that define the Hessian of the energy can be calculated as~\cite{mitarai2020theory}:

\begin{align}
    \frac{\partial^2 E(\textbf{R})}{\partial R_i \partial R_j} = \displaystyle\sum_{a} \Big[ \frac{\partial \theta^{*}_a(\bm{R})}{\partial R_i} \frac{\partial}{\partial \theta_a} \frac{\partial E(\bm{R})}{\partial R_j} \Big]\nonumber + \Big\langle \psi(\theta^{*}(\textbf{R})) \Big| \frac{\partial^2 H(\bm{R})}{\partial R_i \partial R_j} \Big|\psi(\theta^{*}(\bm{R}))\Big\rangle.
\end{align}
To evaluate this expression, it is necessary to compute the expectation value of the Hessian of the Hamiltonian $\partial^2 H(\bm{R})/\partial R_i \partial R_j$, which again can be performed exactly using PennyLane's differentiable Hartree-Fock solver. The need for exact calculations is even more important in this case since approximate techniques such as finite-difference are notoriously problematic for Hessians. \\

To compute $\frac{\partial}{\partial \theta_a} \frac{\partial E(\bm{R})}{\partial R_j}$, it suffices to take the circuit that has been optimized to prepare the ground-state energy and compute its gradient with respect to the expectation vale of $\partial H(\bm{R})/\partial R_j$. This can be performed natively with PennyLane by computing gradients of a circuit that has been optimized to calculate the ground-state energy. The remaining term $\partial \theta^*_a(R)/\partial R_i$ can be obtained by solving the response equations~\cite{mitarai2020theory}:

\beq\label{Eq: response}
    \sum_b \frac{\partial}{\partial \theta_a}\frac{\partial E(\bm{R})}{\partial \theta_b}
    \frac{\partial \theta^*_b(\bm{R})}{\partial R_i}=-\frac{\partial}{\partial \theta_a}
    \frac{\partial E(\bm{R})}{\partial R_j}.
\eeq
We previously discussed how $\frac{\partial}{\partial \theta_a}
    \frac{\partial E(\bm{R})}{\partial R_j}$ can be computed. The term $\frac{\partial}{\partial \theta_a}\frac{\partial E(\bm{R})}{\partial \theta_b}$ is the second-order derivative of the circuit with respect to the expectation value of the Hamiltonian, evaluated at the optimal circuit parameters. It can also be performed natively with PennyLane by expressing the Hessian as the Jacobian of the gradient. Therefore all terms in the equation can be calculated except for
$\partial \theta^*_b(R)/\partial R_i$. This can be obtained by constructing the response equations of Eq.~\eqref{Eq: response} and solving the resulting system of linear equations.\\

Computing energy Hessians provides information about the vibrational structure of molecules, which can be used as inputs to algorithms for calculating properties such as vibronic spectra and vibrational dynamics~\cite{huh2015boson, sparrow2018simulating, jahangiri2020quantum, jahangiri2021quantum}. In the harmonic approximation, the potential experienced by the nuclei is approximated by a quadratic function that leads to harmonic motion. The eigenvectors of the energy Hessian correspond to the normal modes of vibration in the harmonic approximation, and the eigenvalues are the vibrational normal mode frequencies. 

\section{Fully-differentiable workflows}
This section combines concepts described throughout the manuscript to illustrate how PennyLane can be used to create differentiable quantum chemistry workflows for electronic structure calculations capable of simultaneously optimizing circuit parameters, nuclear coordinates, and basis set parameters. The example code below illustrates this for the H$_3^+$ cation.

\begin{code}
	\pythonfile{code/full_optimization.py}
	\caption{Workflow for simultaneously optimizing circuit parameters, nuclear coordinates, and basis set parameters for the H$_3^+$ cation.}
	\label{code:full_optimization}
\end{code}

\section{Qubit tapering}
The number of qubits required to simulate a molecular Hamiltonian can be reduced in the presence of symmetries~\cite{bravyi2017tapering, setia2020reducing}. PennyLane provides functions for tapering off qubits from Hamiltonians containing multiple $\mathbb{Z}_2$ symmetries~\cite{bravyi2017tapering}. The qubit tapering method requires finding a unitary operator that transforms the molecular Hamiltonian to a new Hamiltonian that has the same eigenvalues. The new Hamiltonian always acts trivially, e.g., with an identity or a Pauli-$X$ operator, on a set of qubits $q(j)$. This guarantees that each term of the transformed Hamiltonian commutes with each of the Pauli-$X$ operators applied to the same set of qubits. These commuting terms have the same eigenvectors, which allows removing the qubits from the Hamiltonian terms and replacing them with eigenvalues of the Pauli-$X$ operators $\pm 1$. The unitary operator $U$ is constructed as a Clifford operator given by~\cite{bravyi2017tapering} 
\beq\label{Eq: response}
    U = \Pi_j \left [\frac{1}{\sqrt{2}} \left (X^{q(j)} + \tau_j \right) \right],
\eeq
where $\tau$ denotes the generators of the symmetry group of the Hamiltonian, and the Pauli-$X$ operators act on those qubits $q(j)$ that will be ultimately tapered off from the Hamiltonian. The following code shows how to obtain $\tau$ and $X$ for a given Hamiltonian in PennyLane and use them to taper off qubits. The same symmetries can be used to taper the initial state and the excitation gates to perform a full VQE simulation with a smaller number of qubits.

\begin{code}
	\pythonfile{code/tapering.py}
	\caption{Tapering off qubits in the presence of multiple $\mathbb{Z}_2$ symmetries.}
	\label{code:tapering}
\end{code}

The original Hamiltonian in this code example requires four qubits, two of which can be tapered off using the symmetries.

\section{Resource estimation}
Quantum algorithms such as quantum phase estimation (QPE) and the variational quantum eigensolver (VQE) can be applied to problems that are intractable for conventional computers. These problems require quantum computers capable of implementing large-scale versions of these algorithms, which are not yet available. This makes it difficult to properly explore the accuracy and efficiency of quantum algorithms for problem sizes where the actual advantage of the algorithms can potentially occur. However, it is still possible to estimate the resources required to implement these algorithms for a given problem.

PennyLane has a functionality for estimating quantum resources required to simulate a molecular system. For example, PennyLane can be used to estimate the total number of shots required to estimate the expectation value of an observable~\cite{gonthier2022measurements}. This is relevant for the VQE algorithm, where expectation values are needed to compute gradients and also the final output of the circuit. The following code shows how to estimate the number of shots needed to compute the expectation value of the water Hamiltonian:

\begin{code}
	\pythonfile{code/shots.py}
	\caption{Estimating the number of shots needed to compute the ground state energy with VQE.}
	\label{code:shots}
\end{code}

Resource estimation in PennyLane can also be used to compute the total number of non-Clifford gates and logical qubits required to run quantum phase estimation algorithms for quantum chemistry. The following example shows how to perform such estimation for the water molecule using a double low-rank factorization algorithm~\cite{vonburg2021quantum, lee2021even} for the QPE simulation:

\begin{code}
	\pythonfile{code/qpe_second.py}
	\caption{Estimating the number of non-Clifford gates and logical qubits with the double low-rank factorization algorithm.}
	\label{code:qpe_second}
\end{code}

The total number of non-Clifford gates and logical qubits can also be estimated for first-quantized Hamiltonians of periodic materials in a plane wave basis~\cite{su2021fault}. The cost of these algorithm depends only on the number of plane waves, the volume of the unit cell, and the number of electrons in the material. The following code uses dilithium iron silicate $\text{Li}_2\text{FeSiO}_4$ as an example, whose unit cell consists of 156 electrons and has a volume of $1,145 a_0^3$, where $a_0$ is the Bohr radius~\cite{delgado2022simulating}:

\begin{code}
	\pythonfile{code/qpe_first.py}
	\caption{Estimating the number of non-Clifford gates and logical qubits for a periodic material.}
	\label{code:qpe_first}
\end{code}

\section{Conclusion}
Quantum chemistry is arguably the leading application of quantum computing: it combines (i) a mathematical foundation underpinning the long-term potential for quantum advantage, and (ii) an area of significant industrial interest. Scientists working at the interface between quantum computing and quantum chemistry face the challenge of gaining expertise in these two fields, both of which have a considerable barrier of entry. Software plays a central role in lowering these obstacles and boosting scientific progress. This is one of the main goals of PennyLane: to create a tool that can enhance progress in quantum computing. With this technical manuscript, we aim to provide an additional resource for users that benefit from a deeper dive into the scientific concepts that underly PennyLane's quantum chemistry functionality, enabling them to better employ the software and advance research combining differentiable programming, quantum computing, and quantum chemistry. 

The functionality described in this manuscript reflects PennyLane's capabilities at the time of writing. Our goal is to continue working on further development to expand its scope and enhance its performance. As an open-source software library, PennyLane welcomes contributions from users across the world. This effort will constitute incorporation of state-of-the-art techniques in the scientific literature as well as innovations from the team of PennyLane developers. Quantum computational chemistry is a relatively young field and there still remains much to be discovered.

\bibliography{references}

\end{document}